\documentclass[final,preprint,aps,showpacs,fleqn]{revtex4}
\usepackage{bm}
\usepackage{latexsym,amssymb,amsfonts,amsmath}
\usepackage{graphicx}
\usepackage{setspace}

\newcommand{\bea}{\begin{eqnarray}}
\newcommand{\eea}{\end{eqnarray}}

\newcommand{\be}{\begin{equation}}
\newcommand{\ee}{\end{equation}}
\newcommand{\ba}{\begin{array}}
\newcommand{\ea}{\end{array}}

\begin{document}
\preprint{}
\title {
Dynamical gluon mass at non-zero temperature in instanton vacuum model}
\author{
M. Musakhanov\footnote{E-mail address:
musakhanov@gmail.com },
Sh. Baratov, N. Rakhimov}
\affiliation{
Theoretical  Physics Dept, 
Uzbekistan National University,\\
 Tashkent 100174, Uzbekistan
}
\begin{abstract}

In the framework of the instanton liquid model (ILM), we consider
thermal modifications of the gluon properties in different scenarios
of temperature $T$ dependence of the average instanton size $\bar{\rho}(T)$
and the instanton density $n(T)$ known from the literature. Due to
interactions with instantons, the gluons acquire the dynamical temperature
dependent \char`\"{}electric\char`\"{} gluon mass $M_{el}(q,T).$
We found that at small momenta and zero temperature $M_{el}(0,0)\approx362\,{\rm MeV}$
at the phenomenological values of $\bar{\rho}(0)=1/3\,{\rm fm}$ and
$n(0)=1\,{\rm fm}^{-4}$, however the $T$-dependence of the mass
is very sensitive to the temperature dependence of the instanton vacuum
parameters $\bar{\rho}(T),\,n(T)$: it is very mild in case of the
lattice-motivated dependence and decreases steeply in the whole range
with theoretical parametrization. We see that in region $0<T<T_{c}$
ILM is able to reproduce lattice results for the dynamical  \char`\"{}electric\char`\"{} gluon mass. 
\end{abstract}
\maketitle

\section{Introduction}

The gluodynamics at non-zero temperature $T(\equiv1/\beta)$ is described
by the partition function 
\begin{eqnarray}
Z=\int DA_{\mu}\exp\left\{ -\frac{1}{2g^{2}}\int_{0}^{\beta}dx_{4}\int d^{3}x\,{\rm tr}\,F_{\mu\nu}F_{\mu\nu}\right\} ,
\end{eqnarray}
where $F_{\mu\nu}=\partial_{\mu}A_{\nu}-\partial_{\nu}A_{\mu}-i[A_{\mu},A_{\nu}]$,
and the gauge field $A_{\mu}$ satisfies the periodic condition $A_{\mu}\left(\vec{x},\,x_{4}+\beta\right)=A_{\mu}\left(\vec{x},\,x_{4}\right).$
The extension of the zero-temperature instanton solution \cite{BPST},
caloron, found in \cite{HS} has the form
\begin{eqnarray}
 &  & A_{\mu}^{I}=\Pi\bar{\eta}_{\mu\nu}^{a}(\tau_{a}/2i)\partial_{\nu}\Pi^{-1},\,\,F_{\mu\nu}=\frac{1}{2}\Pi(\tau\partial)\bar{\eta}_{\mu\nu}^{a}(\tau_{a}/2i)(\tau^{+}\partial)\Pi^{-1},\,\,\Pi^{-1}\partial^{2}\Pi=0,\\
 &  & \Pi(r,t)=1+\frac{\pi\rho^{2}}{\beta r}\sinh\frac{2\pi r}{\beta}/\left(\cosh\frac{2\pi r}{\beta}-\cos\frac{2\pi t}{\beta}\right)=1+\sum_{n=-\infty}^{\infty}\frac{\rho^{2}}{r^{2}+(t-n\beta)^{2}}\label{Pi}
\end{eqnarray}
where $r=|\vec{x}|$, $t=x_{4}$ and $\tau_{\mu}=(\vec{\tau},i)$.
At small distances $r,\,t\ll\beta$ the profile $\Pi(x)$ may be approximated
as 
\begin{eqnarray}
\Pi(x)\approx\Pi_{0}(x)=\left(1+\frac{\lambda^{2}}{3}\right)+\rho^{2}/x^{2},\label{Pi0}
\end{eqnarray}
where $\lambda=\pi\rho/\beta$, so the gluon field has an instanton-like
shape with modified instanton size, 
\begin{eqnarray}
A_{\mu}^{I,a}=\frac{2\rho^{'2}}{x^{2}}\,\frac{\bar{\eta}_{\mu\nu}^{a}x_{\nu}}{\left(x^{2}+\rho^{'2}\right)},\,\,\,\rho^{'2}=\rho^{2}/\left(1+\frac{\lambda^{2}}{3}\right).\label{A0}
\end{eqnarray}
In fact, the accuracy of the approximation~~(\ref{Pi0}) is about
one per cent up to $r,\,t\sim\beta$. The extension of the instanton
vacuum liquid model (ILM)~\cite{Diakonov:2002fq,Shuryak1996} to
non-zero temperature in this regime is straightforward and might be
encoded in the temperature dependencies of main parameters of the
model, the average instanton size $\bar{\rho}(T)$ and average instanton
density $n(T)=NT/V_{3}=1/R^{4}(T)$, where $N$ is the total number
of instantons~\cite{DM1988}. Both $\bar{\rho}(T)$ and $n(T)$ in
ILM are homogeneously decreasing functions of $T$.

But the simplified approximation~~(\ref{Pi0}) does not describe
nontrivial phase transition near the critical temperature $T_{c}\sim\Lambda_{QCD}.$
Indeed, for $T<T_{c}$ all color objects are bound into colorless
hadrons. The heat bath predominantly consists of weakly interacting
pions, so the $T$-dependence of instanton density $n(T)$ should
be rather mild, which agrees with expectation of almost constant $T$-dependence
$n(T)=n_{0}(1+O(T^{2}/(6f_{\pi}^{2})))$~\cite{Shuryak1996}. However,
this behavior changes during phase transition, and the expected instanton
density $n(T)$ should be exponentially suppressed at large temperature
$T\gtrsim T_{c}$~\cite{Shuryak1996}.

The extension of ILM which is able to describe the phase transition
from confined to deconfined phase near the critical temperature $T_{c}$
is the so-called dyon-instanton liquid model (DLM)~\cite{1705.04707Shuryak}.
The authors of~\cite{1705.04707Shuryak} concluded later in~\cite{1802.00540Shuryak}
that at very low temperature, the semi-classical description of the
Yang-Mills state reconciles the instanton liquid model without confinement,
with the t'Hooft-Mandelstam proposal of confinement. In the former,
the low temperature thermal state is composed of a liquid of instantons
and anti-instantons, while in the latter it is a superfluid of monopoles
and anti-monopoles.

The temperature dependence of the QCD vacuum model might be tested
by comparison  with results of lattice simulations. For example, for
the ``electric'' gluon mass $M_{el}\left(T\right)$ in the framework
of lattice QCD~\cite{Silva,Maas}, it was found that for $T\geq T_{c}$
the linear dependence $M_{el}(T)\sim T$ is consistent with Debye
screening and has a minimum at $T\sim T_{c}$, whereas for $T\leq T_{c}$
the electric mass $M_{el}(T)$ is a rather slowly decreasing function
of temperature $T$. This behavior might be naturally explained in
the framework of the ILM, which predicts the temperature dependence
like 
\begin{equation}
M_{el}\sim({\rm packing\,\,\,parameter}(T))^{1/2}\bar{\rho}^{-1}(T)=\bar{\rho}(T)n^{1/2}(T),
\end{equation}
 a decreasing function of temperature at $T\leq T_{c}$. Combined
with perturbative one-loop thermal gluon contribution to the gluon
propagator, which is rising with temperature as $M_{pert,el}(T)\sim T$,
this model is able to reproduce lattice results for the dynamical
gluon mass~\cite{Silva,Maas}.

There are two major technical challenges in calculation of the gluon
propagator in ILM framework: the account of zero-modes (fluctuations
along of instanton collective coordinates), and the averaging over
the collective coordinates of all instantons. We address the former
using the approach of~\cite{Brown}, while for the latter we extend
Pobylitsa's approach~\cite{Pobylitsa}, applied earlier by us for
the gluons at $T=0$~\cite{MO2018}, and consider in this paper its
further extension for the ILM averaged gluon propagator at $T\neq0$.

The paper is structured as follows. In the Section~\ref{sec:variation}
we review briefly the formulation of ILM at nonzero temperature $T\not=0$
and discuss the temperature dependence of main instanton vacuum parameters.
In Section~\ref{sec:Scalar} we consider a simplified case and evaluate
the propagator of scalar color octet particle (which we call ``scalar
gluon'') in the instanton background at nonzero temperature. It allows
us to get several important results which will be used later. In the
Section~\ref{sec:Real} we consider the case of real gluon and evaluate
the propagator at nonzero temperature. We extract the electric mass
$M_{el}$ and compare it with lattice results. Finally in Section~\ref{sec:Conclusion}
we draw conclusions.

\section{Variational estimates in ILM at $T\protect\neq0$}

\label{sec:variation}
\begin{figure}[h]
\includegraphics[scale=0.6]{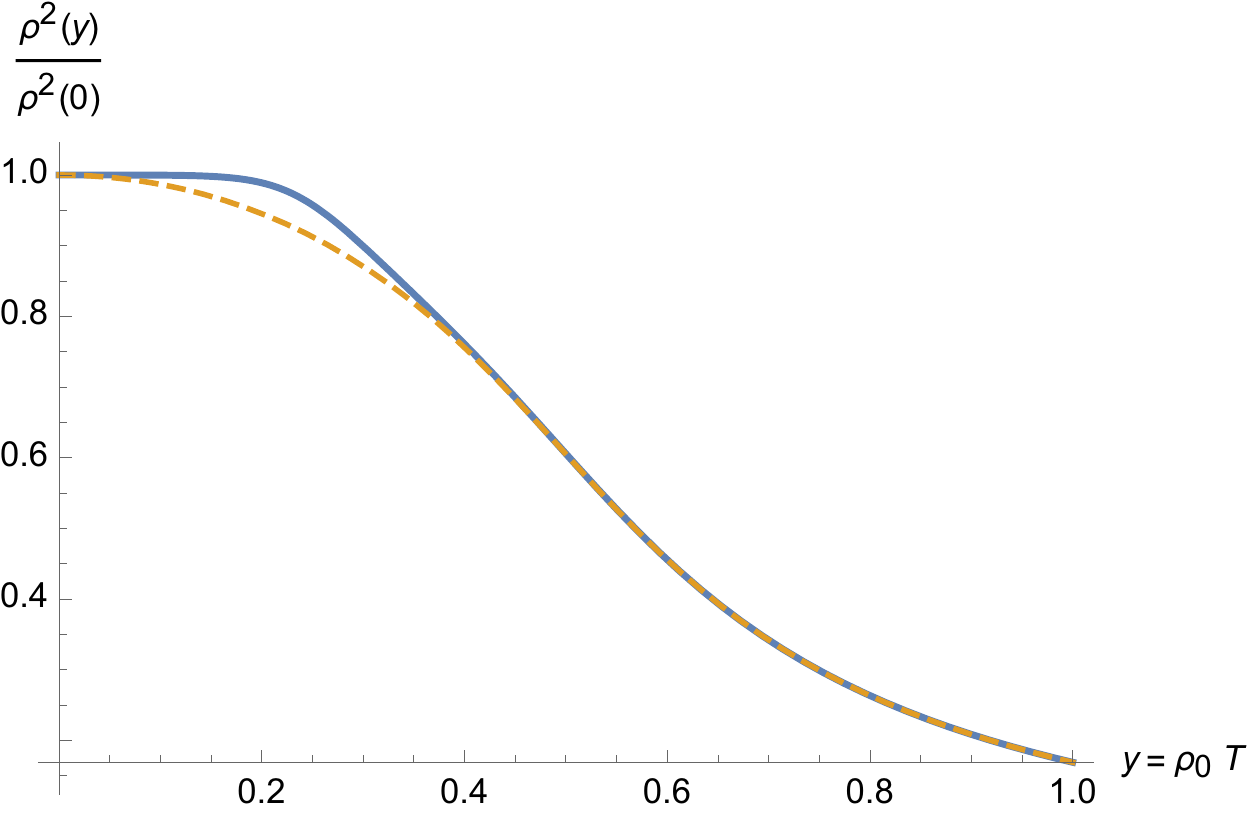}
\includegraphics[scale=0.6]{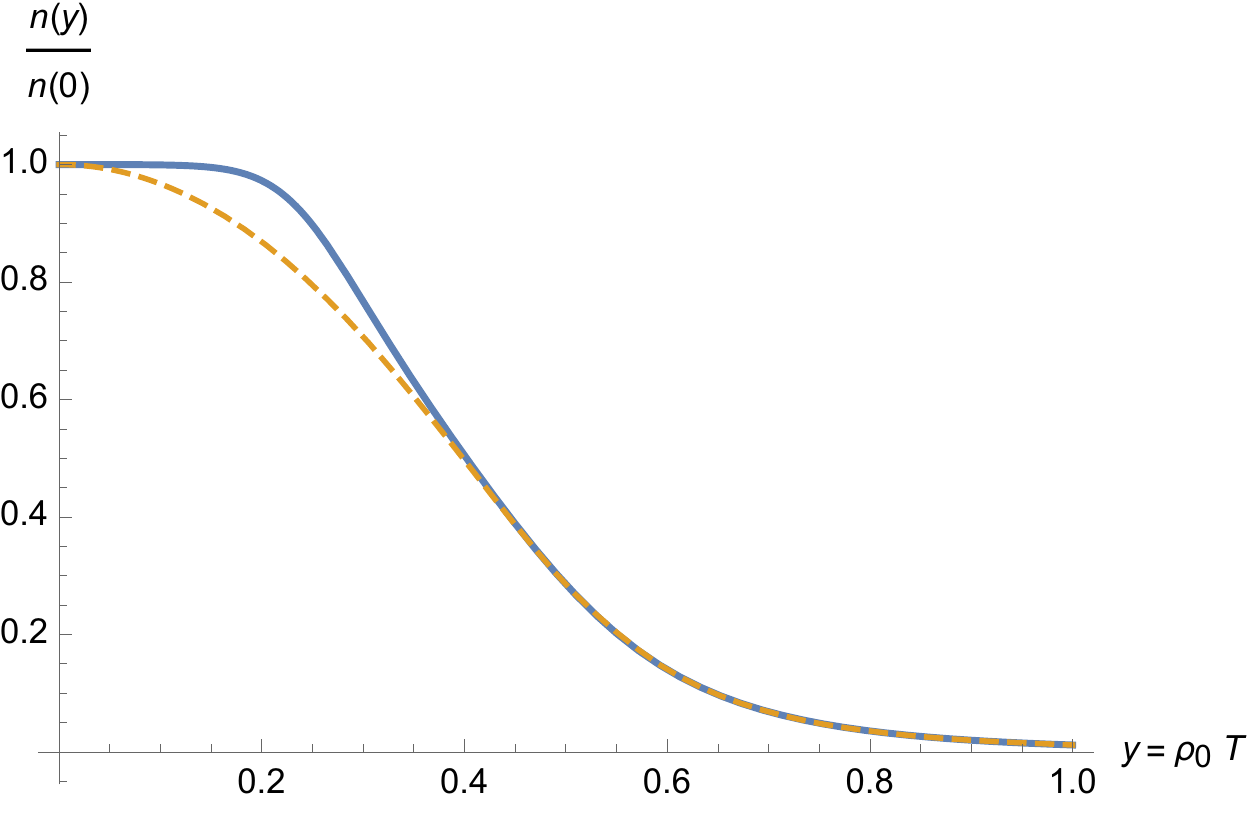}
\caption{The (normalized to unity) temperature dependence of the instanton
size ${\bar{\rho}}^{2}(y)/{\bar{\rho}}^{2}(0)$ (left plot ) and (normalized
to unity) temperature dependence of instanton density $n(y)/n(0)$
(right plot, obtained with variational estimates from Refs.~\cite{DM1988,Shuryak1996,SeungilNam}).
In both plots we use $y=\bar{\rho}_{0}T$ notations and phenomenological
values  $\bar{\rho}(0)=1/3\,{\rm fm}$ and $n(0)=1\,{\rm fm}^{-4}$
for estimates. The solid line corresponds to evaluation with the modification
$A_{N_{c}}\rightarrow A_{N_{c}}\Theta_{\Delta y}(y-y_{c})$ in~~(\ref{Phi}),
where $\Theta_{\Delta y}(y-y_{c})$ is a smooth interpolating step-like
function Eq.(\ref{theta})  with a width $\Delta T=0.3T_{c}$~\cite{Shuryak1996}.
Dashed lines correspond to the evaluation with $A_{N_{c}}={\rm const}$
in (\ref{Phi}) (full suppression) for all $T$.}
\label{Shuryak1996} 
\end{figure}

The application of the Feynman variational principle
to the QCD vacuum filled with instanton gas leads to the ILM~\cite{DP1984},
which was  generalized to non-zero temperatures in~\cite{DM1988}.
Main variational ingredients of this approach are the instanton size
distribution function $\mu(\rho,\,T,\,n)$ and the density of instantons
$n(T)$~\cite{DM1988,Shuryak1996,SeungilNam}. The ILM instanton
size distribution function is closely related to the thermal single
instanton one-loop distribution function~\cite{gross1981} 
\begin{equation}
d(\rho,T)=C\rho^{b-5}\exp(-A_{N_{c}}T^{2}),\qquad C={\rm const}
\end{equation}
, 
\begin{eqnarray}
 &  & \mu(\rho,\,T,\,n)=C\,\rho^{b-5}\exp\left[-\Phi(n,T)\rho^{2}\right],\label{mu}\\
 &  & \Phi(n,\,T)=\frac{1}{2}A_{N_{c}}\,T^{2}+\left[\frac{1}{4}A_{N_{c}}^{2}T^{4}+\nu\bar{\beta}\gamma^{2}n\right]^{1/2}.\label{Phi}
\end{eqnarray}
where $A_{N_{c}}=1/3\,[11/6\,N_{c}-1]\pi^{2}$, $b=11/3\,N_{c}$,
$\nu=\frac{b-4}{2}$, $\bar{\beta}=-b\log(\Lambda\bar{\rho})$, $\gamma^{2}=\frac{27\pi^{2}N_{c}}{4(N_{c}^{2}-1)}$,
and 
\begin{eqnarray}
{\bar{\rho}}^{2}(T,\,n)=\frac{1}{\mu_{0}(T,n)}\int_{0}^{\infty}d\rho\mu(\rho,T,\,n)\rho^{2},\,\,\,\mu_{0}(T,\,n)=\int_{0}^{\infty}d\rho\mu(\rho,T,\,n).
\end{eqnarray}
The result~(\ref{mu}) might be obtained maximizing the  variational
ILM partition function~\cite{DM1988}  
\begin{eqnarray}
 &  & Z\geq Z_{1}\left(\mu,\,n\right)\exp(-\langle E-E_{1}\rangle),\label{eq:Z0}\\
 &  & Z_{1}\left(\mu,\,n\right)=\frac{1}{(N/2\,!)^{2}}\left(\frac{2\mu_{0}(T,n)V}{N}\right)^{N},\,\,\,\langle E-E_{1}\rangle=-\frac{N}{2}\bar{\beta}\gamma^{2}n{\bar{\rho^{2}}}^{2}\label{Z}
\end{eqnarray}
with respect to parameter $\mu$.The minimization of the free energy
$F=-T/V_{3}\log Z$ ~ by variation over $n$ leads to the equation
for the density 
\begin{eqnarray}
n(T)=2\mu_{0}(n,T)
\end{eqnarray}
The variational estimates demonstrated that $\bar{\rho}(T)$ and $n(T)$
are decreasing functions of temperature $T$~(see fig.~\ref{Shuryak1996})
due to the exponential factor in $d(\rho,T)\sim A_{N_{c}}$. On the
other hand, lattice data show that the instanton density $n$ is not
modified by temperature up to critical temperature $T_{c}$~\cite{Chu1995}.
In numerical simulations of ILM~\cite{Shuryak1996} it was suggested
to interpolate between no suppression ($A_{N_{c}}=0$ in Eq.~(\ref{Phi}))
below $T_{c}$ and full suppression ($A_{N_{c}}\neq0$ in Eq.~(\ref{Phi}))
above $T_{c}\sim150\,{\rm MeV}$ , with a width $\Delta T=0.3T_{c}$
to be in the correspondence with lattice result~\cite{Chu1995}.
We are following this suggestion and are repeating the calculations
with the modification in the Eq.~(\ref{Phi}) as $A_{N_{c}}\rightarrow A_{N_{c}}\Theta_{\Delta y}(y-y_{c}),$
where 
 $y=\bar{\rho}_{0}T$, $y_{c}=\bar{\rho}_{0}T_{c}=0.25\sim T_{c}=150\,{\rm MeV},\,\,\,\Delta y=\bar{\rho}_{0}\Delta T_{c}=0.075\sim\Delta T=0.3T_{c}$,
and we introduced a smooth interpolation with a step-like function
of width $\Delta y$,
\begin{eqnarray}
\Theta_{\Delta y}(z)=1/2\,[1+\tanh(z/\Delta y)].\label{theta}
\end{eqnarray}
The temperature dependence of $\bar{\rho^{2}}(y)/\bar{\rho^{2}}(0)$
and $n(y)/n(0)$ is shown in the  Fig.~~(\ref{Shuryak1996}). 

\section{Color octet scalar propagator at non-zero temperature}

\label{sec:Scalar}

We start from the scalar massless field $\phi$ belonging to the adjoint
representation, the same as a physical gluon. We have to find its
propagator in the external classical gluon field  of instanton gas
$A_{\mu}=\sum_{I}A_{\mu}^{I}(\gamma_{I})$, where $A_{\mu}^{I}(\gamma_{I})$
is a generic notation for the QCD (anti-) instanton, and $\gamma_{I}$
stands for all the relevant collective coordinates: the position in
Euclidean 4D space $z_{I}$, the size $\rho_{I}$ and the $SU(N_{c})$
color orientation $U_{I}$ ($4N_{c}$ collective coordinates in total).
The averaging over the instanton collective coordinates includes the
integral over the instanton position $\int d^{4}z\equiv\int_{V_{3}}d^{3}z_{I}\int_{0}^{\beta}dz_{I,4}$.
In view of periodicity of the fields $\phi(\vec{x},t+\beta)=\phi(\vec{x},t)$
at nonzero temperature, we may restrict the integration over $t$
to the period $\beta$, so the effective action takes a form 
\begin{eqnarray}
S_{\phi}=\int_{V_{3}}d^{3}x\int_{0}^{\beta}dt\,\phi^{\dagger}(\vec{x},t)P^{2}\phi(\vec{x},t)\label{Sphi}
\end{eqnarray}
where $P_{\mu}=p_{\mu}+A_{\mu}$ (in the coordinate representation
$p_{\mu}=i\partial_{\mu}$). The color octet scalar
propagator in the field of instanton gas is given by 
\begin{eqnarray}
 &  & \Delta=(p+A)^{-2}=(p^{2}+\sum_{i}(\{p,A_{i}\}+A_{i}^{2})+\sum_{i\neq j}A_{i}A_{j})^{-1},\,\,\,\Delta_{0}=p^{-2},\label{tildeDelta}
\end{eqnarray}
It is convenient to introduce also the propagators ``gluon'' in
the field of individual instantons,
\begin{equation}
\Delta_{i}=P_{i}^{-2}=(p^{2}+\{p,A_{i}\}+A_{i}^{2})^{-1},
\end{equation}
and in the instanton gas background when the overlaps $\sim\sum_{i\neq j}A_{i}A_{j}$
are disregarded, 
\begin{equation}
\tilde{\Delta}=(p^{2}+\sum_{i}(\{p,A_{i}\}+A_{i}^{2}))^{-1},
\end{equation}

There are no zero modes in $\Delta_{i}^{-1}=P_{i}^{2}$ nor in $\Delta^{-1}=P^{2}$,
which means the existence of the inverse operators $\Delta_{i}$ and
$\Delta$. Our aim is to find the propagator averaged over instanton
collective coordinates $\bar{\Delta}\equiv\langle\Delta\rangle=\int D\gamma\,\Delta.$
In coordinate space the propagator $\Delta$ and free propagator $\Delta_{0}$
must be periodical functions of time with period $\beta$. In what
follows we will use notation $\Delta(x,\,x')\equiv\langle x|\Delta|x'\rangle$
for the matrix element of the operator $\Delta$ between Fock states
 labeled by space-time coordinates $|x\rangle\equiv|\vec{x},t\rangle=|\vec{x}\rangle|t\rangle$
(the same for $|x'\rangle$) and $\left[x_{\mu},\,p_{\nu}\right]=i\delta_{\mu\nu}$~\cite{Schwinger1951}.
These states form a complete orthonormalized set, $\sum_{t}|t\rangle\langle t|=1,\,\,\langle t'|t\rangle=\delta(t'-t).$
Also, we define the step-operator, $\langle t'|\Theta|t\rangle=\Theta(t'-t)$.
In view of $\frac{d}{dt'}\Theta(t'-t)=\delta(t'-t)$, we may conclude
that $\Theta^{-1}\equiv\frac{d}{dt}$. The time-periodic state with
period $\beta$ may be represented in terms of states $|t\rangle$
as 
\begin{eqnarray}
\left|t_{\beta}\right\rangle \equiv\sum_{n=-\infty}^{\infty}\left|t-n\beta\right\rangle ,\,\,\,\left\langle t'|t_{\beta}\right\rangle =\delta\left(t'-t_{\beta}\right)\equiv\sum_{n=-\infty}^{\infty}\delta\left(t'-(t-n\beta)\right).\label{tbeta}
\end{eqnarray}
Now the evaluation of the propagator $\Delta=\left(P^{2}\right)^{-1}$ is
straightforward. 
Since $\left\langle t'\left|P^{2}\right|t_{\beta}\right\rangle =\delta(t'-t_{\beta})({\vec{P}}^{2}+(i\frac{\partial}{\partial t'}+A_{4})^{2})$,
we have the equation in the form 
\begin{eqnarray}
({\vec{P}}^{2}+(i\frac{\partial}{\partial t'}+A_{4})^{2})\langle t'|\Delta|t_{\beta}\rangle=\delta(t'-t_{\beta})
\nonumber\\
\Rightarrow\langle\vec{x}',t'|\Delta|\vec{x},t_{\beta}\rangle
=\sum_{n=-\infty}^{\infty}\Delta(\vec{x}',t'|\vec{x},t-n\beta),
\end{eqnarray}
where $\Delta(\vec{x}',t'|\vec{x},t-n\beta)$ is a usual zero-temperature
($T=0$) aperiodical propagator. For physical applications we need
to average the propagator $\Delta$  over  collective coordinates
of all instantons $\,\,\,\,\,\bar{\Delta}\,\,=\,\,\langle\Delta\rangle\,=\,\int D\gamma\,\Delta.$
We are following the procedure developed in our previous paper~\cite{MO2018},
where the approach~\cite{Pobylitsa} derived for the quark correlators,
was extended to the gluon case. We start first from averaging over
collective coordinates of the operator$\bar{\tilde{\Delta}}=\langle\tilde{\Delta}\rangle$
(see Eq.~~(\ref{tildeDelta})). Since Pobylitsa Eqs.~\cite{Pobylitsa,MO2018}
are written in operator form, they can be easily extended to $T\ne0$
case just by calculating of matrix elements of propagator $\tilde{\Delta}$
with periodical states $|t_{\beta}\rangle$ on the right side.

Since the instanton gas is dilute, namely the packing parameter $\rho^{4}n\sim(1/3)^{4}=1.2\times10^{-1}\ll1$,
we may develop a systematic expansion over the parameter $n$. The
expansion of the inverse propagator up the first-order $\mathcal{O}(n)$
terms has a form 
\begin{eqnarray}
\bar{\tilde{\Delta}}^{-1}-\Delta_{0}^{-1}=\left\langle \sum_{i}\{\Delta_{0}+(\Delta_{i}^{-1}-\Delta_{0}^{-1})^{-1}\}^{-1}\right\rangle =N\,\Delta_{0}^{-1}(\bar{\Delta}_{I}-\Delta_{0})\Delta_{0}^{-1}+\mathcal{O}(n^{2}),\label{Eqdelta}
\end{eqnarray}
where $\bar{\Delta}_{I}=\int d\gamma_{I}\,\Delta_{I}$ is the propagator
in the field of individual instanton averaged over its collective
degrees of freedom. In the same order of expansion we may approximate
the inverse propagator as $\bar{\Delta}^{-1}=\bar{\tilde{\Delta}}^{-1}=p^{2}+M_{s}^{2}$,
where we introduced squared dynamical color octet scalar
mass operator $M_{s}^{2}$ whose matrix elements are given by 
\begin{eqnarray}
\left\langle t'|M_{s}^{2}|t_{\beta}\right\rangle \delta_{ab}=N\,p^{2}(\langle t'|\bar{\Delta}_{I}^{ab}|t_{\beta}\rangle-\langle t'|\Delta_{0}^{ab}|t_{\beta}\rangle)p^{2}.\label{Ms}
\end{eqnarray}
According to \cite{gross1981}, the periodic color octet scalar
propagator in instanton field (\ref{Pi-}) is given by 
\begin{eqnarray}
 &  & \Delta_{I}^{ab}(x,y)=\Delta_{0}^{ab}(x,y)+\Delta_{1}^{ab}(x,y)+\Delta_{2}^{ab}(x,y),\label{eq:DeltaFull}\\
 &  & \Delta_{0}^{ab}(x,y)=\frac{1}{2}\,{\rm tr}\left(\frac{\tau_{a}F(x,y)\tau_{b}F(y,x)}{\Pi(x)4\pi^{2}(x-y)^{2}\Pi(y)}\right),\\
 &  & F(x,y)=1+\sum_{m}\frac{\rho^{2}(\tau x_{m})(\tau^{\dagger}y_{m})}{x_{m}^{2}y_{m}^{2}},\,\,\,(x_{m}\equiv x-m\beta\hat{t},\,\,y_{m}\equiv y-m\beta\hat{t})\\
 &  & \Delta_{1}^{ab}(x,y)=\frac{1}{2}\,{\rm tr}\left({\sum_{m}}^{'}\frac{\tau_{a}F(x,y_{m})\tau_{b}F(y_{m},x)}{\Pi(x)4\pi^{2}(x-y_{m})^{2}\Pi(y)}\right)\\
 &  & \Delta_{2}^{ab}(x,y)=\sum_{m}\frac{C^{ab}(x,y_{m})}{\Pi(x)4\pi^{2}\Pi(y)},\label{eq:Delta2}\\
 &  & C^{ab}(x,y)=\sum_{r\neq s}\frac{2\Phi_{rs}^{a}(x)\Phi_{rs}^{b}(y)}{\beta^{2}(r-s)^{2}}-\sum_{r\neq s}\sum_{t\neq u}\frac{\rho^{2}\Phi_{rs}^{a}(x)}{\beta^{2}(r-s)^{2}}\frac{\Phi_{tu}^{b}(y)}{\beta^{2}(t-u)^{2}}h_{rs,tu},\,\,\label{eq:Cab}\\
 &  & \Phi_{rs}^{a}(x)=\frac{\rho^{2}\beta(r-s)x^{a}}{x_{r}^{2}x_{s}^{2}}\label{eq:Phiab}\\
 &  & \sum_{m}{C^{ab}(x,y_{m})}=\sum_{m}\sum_{r\neq s}\frac{2\Phi_{rs}^{a}(x)\Phi_{rs}^{b}(y_{m})}{\beta^{2}(r-s)^{2}}=\sum_{r\neq s}\frac{\rho^{2}x^{a}}{x_{r}^{2}x_{s}^{2}}\sum_{m}\frac{\rho^{2}y^{b}}{y_{r+m}^{2}y_{s+m}^{2}}\nonumber 
\end{eqnarray}
Combining~(\ref{eq:Cab},\ref{eq:Phiab}) we may simplify~(\ref{eq:Delta2})
to
\begin{eqnarray}
\Delta_{2}^{ab}(x,y)=\sum_{r\neq s}\frac{\rho^{2}x^{a}}{x_{r}^{2}x_{s}^{2}}\sum_{m}\frac{\rho^{2}y^{b}}{y_{r+m}^{2}y_{s+m}^{2}}\frac{1}{\Pi(x)4\pi^{2}\Pi(y)}.
\end{eqnarray}
At short distances $r\sim t\leq\beta$ the caloron field becomes instanton-like~~(\ref{A0})
with modified  instanton radius  $\rho^{'2}=\rho^{2}/(1+1/3\,\lambda^{2}),$
and $\lambda=\pi\rho/\beta$. In this region we may simplify the first
term in~(\ref{eq:DeltaFull}) as
\begin{eqnarray}
 &  & \Delta_{I,0}^{ab}=\frac{1}{2}{\rm tr}\left(\frac{\tau_{a}F_{0}(x,y)\tau_{b}F_{0}(y,x)}{4\pi^{2}(x-y)^{2}\Pi_{0}(x)\Pi_{0}(y)}\right),\,\,\,\,\Pi_{0}(x)=\frac{x^{2}+\rho^{'2}}{x^{2}},\label{Delta}\\
 &  & \tau_{\mu}\equiv(\vec{\tau},i),\,\,\,\tau_{\mu}^{\dagger}=(\vec{\tau},-i),\,\,\,\tau_{\mu}\tau_{\nu}^{+}=\delta_{\mu\nu}+i\bar{\eta}_{a\mu\nu}\tau_{a},\\
 &  & F_{0}(x,y)=1+\rho^{'2}\frac{(\tau x)(\tau^{+}y)}{x^{2}y^{2}}=1+\rho^{'2}\frac{(xy)}{x^{2}y^{2}}+\rho^{'2}\frac{i\bar{\eta}_{a\mu\nu}\tau_{a}x_{\mu}y_{\nu}}{x^{2}y^{2}},
\end{eqnarray}
where $\bar{\eta}_{a\mu\nu}=-\bar{\eta}_{a\nu\mu}$ is the 'tHooft
symbol. %The following calculations are quite similar our previous work
%and we may refer to~\cite{MO2018} for the details. 
As will be shown below, the contribution of the terms~$\Delta_{1}^{ab}(x,y),\,\Delta_{2}^{ab}(x,y)$
in~(\ref{eq:DeltaFull}) is small and might be neglected.

The collective degrees of freedom (center of instanton position $z$
and orientation $U$) in~(\ref{Delta}) might be introduced shifting
the arguments $x\rightarrow x-z$, $y\rightarrow y-z$ and color rotation
factors $\Delta_{I}^{ab}\to O^{ab}O^{a'b'}\Delta_{I}^{bb'}$, where
$O^{ab}$ are the matrices of color rotation in adjoint representation
%~\footnote
({
%The color transformation matrices 
$O^{ab}$ are related to color rotation
matrices in fundamental representation by $O^{ab}={\rm tr}\left(U^{\dagger}t^{a}Ut^{b}\right)$,
where $t_{a}$ are $SU(N_{c})$- matrices.}). 
The averaging over the collective coordinates reduces to integration
over the instanton center, $\int_{0}^{\beta}dz_{4}\int_{V_{3}}d^{3}z$,
and color orientation, $\int dO$. The latter integral might be evaluated
analytically using  the well-known identities~\cite{Shuryak1996}:
$
%\begin{align}
\int dOO^{ab}O^{ab'}  =\delta_{bb'},
%\\
\int dOO^{ab}O^{a'b'}  =(N_{c}^{2}-1)^{-1}\delta_{aa'}\delta_{bb'},
%\end{align}
$
$
%\begin{equation}
\int dOO^{ab}\bar{\eta}_{b\mu\nu}O^{a'b'}\bar{\eta}_{b'\mu'\nu'}=(N_{c}^{2}-1)^{-1}\delta_{aa'}(\delta_{\mu\mu'}\delta_{\nu\nu'}-\delta_{\mu\nu'}\delta_{\nu\mu'}).
%\end{equation}
$
The contribution of the $\Delta_{I,0}^{ab}$ to the color octet scalar
dynamical mass operator~(\ref{Ms}) is given by 
\begin{eqnarray}
M_{s,0}^{2}\delta_{ab}=N\,p^{2}\left(\bar{\Delta}_{I,0}^{ab}-\Delta_{0}^{ab}\right)p^{2},\label{Ms0}
\end{eqnarray}
so for the expression in parenthesis in~(\ref{Ms0}) we may obtain
after collective coordinate averaging in the coordinate representation
\begin{eqnarray}
 &  & \bar{\Delta}_{I,0}^{aa'}(x,\,y)-\Delta_{0}^{aa'}(x,\,y)\nonumber \\
 &  & =\int d^{4}z\,dO\,O^{ac}O^{a'c'}\left(\Delta_{I,0}^{cc'}(x',y')-\Delta_{0}^{cc'}(x',y')\right)\,\,\,(x'\equiv x-z,\,\,\,y'\equiv y-z),\nonumber \\
 &  & =\delta_{aa'}\int d^{4}z\left[\frac{3\rho^{'2}}{4\pi^{2}(N_{c}^{2}-1)}f_{1}(x')f_{1}(y')+\frac{2\rho^{'4}}{N_{c}^{2}-1}f_{2}(x')g(x'-y')f_{2}(y')\right],\label{barDelta}
\end{eqnarray}
where we introduced notations
\begin{eqnarray*}
 &  & f_{1}(x)=\frac{1}{\left(x^{2}+\rho^{'2}\right)},\,\,\,f_{2}(x)=\frac{(x_{\mu}x_{\nu},ix^{2})}{x^{2}\left(x^{2}+\rho^{'2}\right)},\,\,\,\,g(x-y)=\frac{1}{4\pi^{2}(x-y)^{2}}.
\end{eqnarray*}
In what follows we will use notation $M_{s}^{2}(\vec{q},\,m)$ for
the dynamical color octet scalar mass  corresponding
to the   Matsubara mode $m$ with frequency $\omega_{m}=2\pi mT$
in the 3-momentum $\vec{q}$ representation.
We are especially interesting with the $m=0$ Matsubara mode, $M_{s}^{2}\left(\vec{q},\,m=0\right)\equiv M_{s}^{2}(\vec{q},\,T)$.

If we define the Fourier transformation to coordinate space as 
\begin{eqnarray}
 &  & f_{1}(x-z)=\sum_{m=-\infty}^{\infty}\int\frac{d^{3}p}{(2\pi)^{3}}\exp[i\vec{p}(\vec{x}-\vec{z})]\exp(2\pi m\,(x-z)_{4}/\beta)f_{1}(\vec{p},\,m),\label{f1}
\end{eqnarray}
then, the contribution of the first term in Eq.~~(\ref{barDelta})
might be rewritten as
\begin{equation}
M_{s,0,1}^{2}(q,T)\sim q^{4}f_{1}(\vec{q},\,m=0)f_{1}(-\vec{q},\,m=0)=q^{4}f_{1}^{2}(\vec{q},\,0),
\end{equation}
where 
\begin{eqnarray}
 &  & q^{2}f_{1}(q,\,0)=q^{2}{\rho'}^{2}\int_{-\beta/2\rho'}^{\beta/2\rho'}dx_{4}\int_{-\infty}^{\infty}dx_{2}dx_{3}\pi\exp\left(-q\rho'\left(\sum_{i=2}^{4}{x_{i}}^{2}+1)^{1/2}\right)\right)\frac{1}{\left(\sum_{i=2}^{4}{x_{i}}^{2}+1\right)^{1/2}}\nonumber \\
 &  & \le q^{2}{\rho'}^{2}\int_{-\infty}^{\infty}dx_{4}dx_{2}dx_{3}\frac{\pi\,\exp\left(-q\rho'(\sum_{i=2}^{4}{x_{i}}^{2}+1)^{1/2})\right)}{\left(\sum_{i=2}^{4}{x_{i}}^{2}+1\right)^{1/2}}=4\pi^{2}q\rho'K_{1}(q\rho'),
\end{eqnarray}
and $K_{1}(z)$ is a modified Bessel function of the second kind,
with $\lim_{z\rightarrow0}z\,K_{1}(z)=1.$ Since temperature mildly
affects dynamical mass form-factor, we may neglect this modification
at small temperatures $T\leq T_{c}$. Careful analysis shows that
second term in Eq.~~(\ref{barDelta}) and all of other terms including
$\Delta_{1}^{ab}$ and $\Delta_{2}^{ab}$ give zero or negligible
contribution, so we finally obtain 
\begin{eqnarray}
M_{s}(q,T)\approx M_{s,0,1}(q,T)=\left[\frac{3\bar{\rho}^{'2}(T)n(T)}{(N_{c}^{2}-1)}4\pi^{2}\right]^{1/2}F(q,T),\label{M01}
\end{eqnarray}
where
\begin{eqnarray*}
F(0,0)=1,\,\,F(q,T)\le F(q,0)=q\bar{\rho}K_{1}(q\bar{\rho}).
\end{eqnarray*}

\section{Gluon propagator at non-zero temperature}

\label{sec:Real}In this section we will extend the calculations of
averaged full gluon propagator $\bar{S}_{\mu\nu}$, considered in~\cite{MO2018},
to non-zero temperature case. It is a rather straightforward task,
since there all principal equations and their solutions were found
in operator form. First, we have the solution of Pobylitsa equation
in operator form as 
\begin{eqnarray}
\bar{S}_{\rho\nu}-S_{\rho\nu}^{0}=N\,\left(\bar{S}_{\rho\nu}^{I}-{S^{0}}_{\rho\nu}\right)+\mathcal{O}\left(n^{2}\right),\label{barS}
\end{eqnarray}
where free and single instanton gluon propagators \cite{BrownCarlitz}
are given by 
\begin{eqnarray}
 &  & {S^{0}}_{\mu\nu}=\left(\delta_{\mu\nu}-(1-\xi)p_{\mu}p_{\nu}/p^{2}\right)/p^{2},\,{S^{0}}_{\mu\nu}^{-1}=\delta_{\mu\nu}p^{2}-(1-1/\xi)p_{\mu}p_{\nu}\label{S0}\\
 &  & S_{\mu\nu}^{I}=q_{\mu\nu\rho\sigma}P_{\rho}^{I}\Delta_{I}^{2}P_{\sigma}^{I}-(1-\xi)P_{\mu}^{I}\Delta_{I}^{2}P_{\nu}^{I},\label{SI}
\end{eqnarray}
and we introduced notation $q_{\mu\nu\rho\sigma}=\delta_{\mu\nu}\delta_{\rho\sigma}+\delta_{\mu\rho}\delta_{\nu\sigma}-\delta_{\mu\sigma}\delta_{\nu\rho}+\epsilon_{\mu\nu\rho\sigma}$
(for the antiinstanton case $+\epsilon_{\mu\nu\rho\sigma}\Rightarrow-\epsilon_{\mu\nu\rho\sigma}$).
The Eq.~~(\ref{barS}) can be rewritten (compare with the Eq.~~(\ref{Eqdelta}))
as 
\begin{eqnarray}
\Pi_{\rho\nu}\equiv\bar{S}_{\rho\nu}^{-1}-S_{0,\rho\nu}^{-1}=N{S^{0}}_{\rho\sigma}^{-1}(\bar{S}_{\sigma\mu}^{I}-{S^{0}}_{\sigma\mu}){S^{0}}_{\mu\alpha}^{-1}+\mathcal{O}(n^{2}).\label{barS-}
\end{eqnarray}
At non-zero temperature, the most essential point is the lack of the
relativistic covariance, since Euclidian time is restricted to the
interval $0\le x_{4}\le\beta=1/T$, and all of the bosonic fields
( background $A_{\mu}$, the fluctuations $a_{\mu}$ and zero-modes
$\phi_{\mu}$) must be time-periodic functions with period $\beta,$
$A_{\mu}(\vec{x},\,x_{4}+\beta)=A_{\mu}(\vec{x},\,x_{4}).$ The operator
form of the main equation~(\ref{barS-}) significantly simplifies
our problem, since at the end we have just to calculate the matrix
element of the operators between time state $|t'\rangle$ and periodical
state $|t_{\beta}\rangle$ defined in Eq~~(\ref{tbeta}). Then by
means of Fourier transformations we can obtain momentum representation
of the propagators written in terms of three-momenta $\vec{k}$ and
Matsubara modes $m$ ($k_{4}=2\pi mT$, $k_{\mu}\equiv\left(\vec{k},\,k_{4}\right)$,
$k^{2}={\vec{k}}^{2}+k_{4}^{2}$).

\begin{figure}[h]
\includegraphics[scale=0.6]{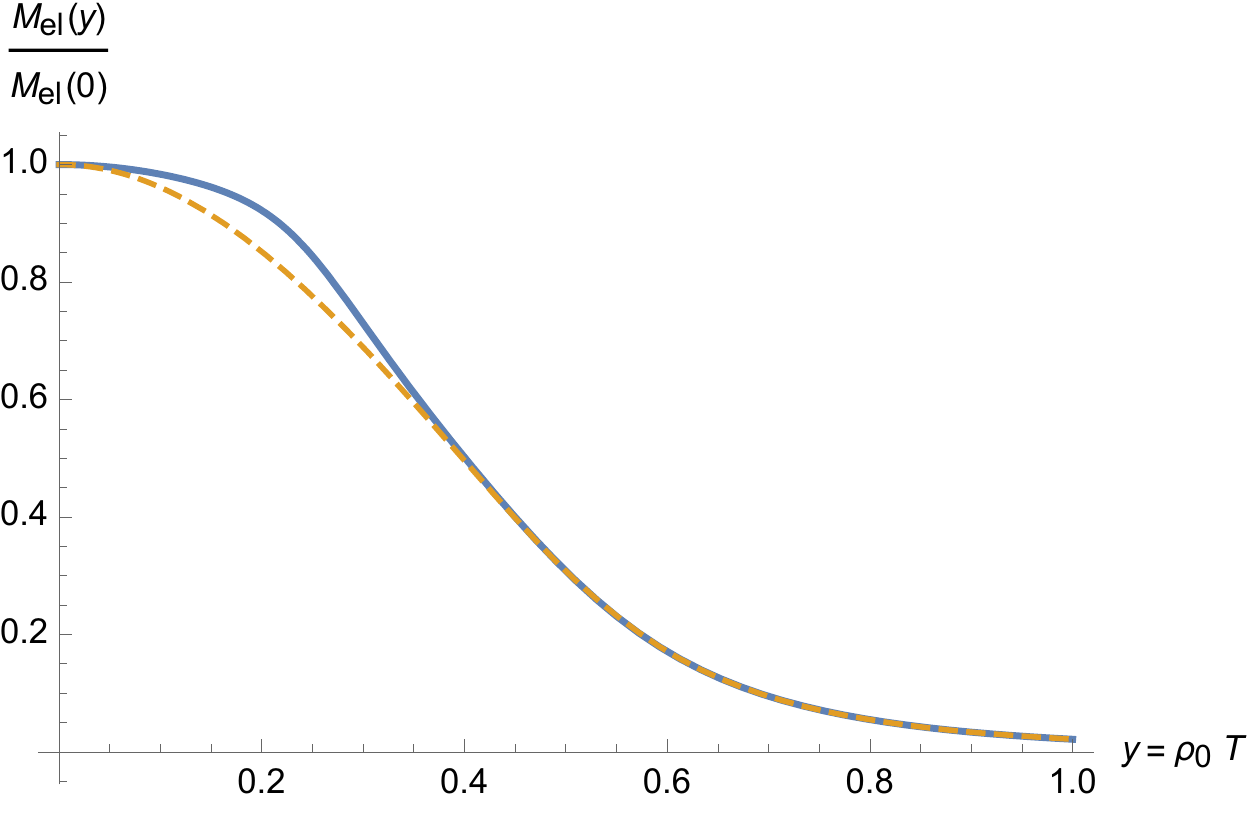} \includegraphics[scale=0.6]{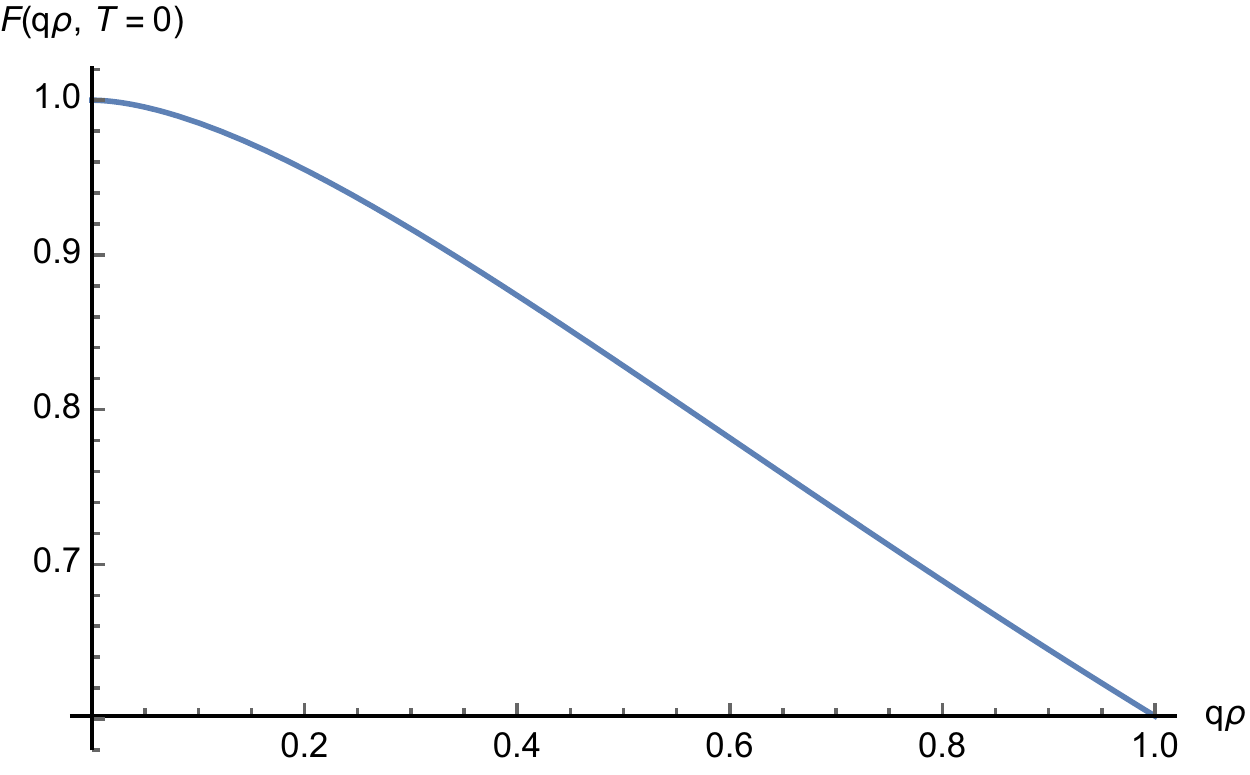}
\caption{Left plot: temperature dependence of \char`\"{}electric\char`\"{}
gluon dynamical mass $M_{el}(0,T)/M_{el}(0,0)$. The solid line was
evaluated using with modification $A_{N_{c}}\rightarrow A_{N_{c}}\Theta_{\Delta y}(y-y_{c})$
(see Eqs.(\ref{Phi},~\ref{theta})) and interpolates smoothly between
no suppression below  critical temperature $T_{c}=150\,\,{\rm MeV}$
and full suppression above it, with a width of the interpolating region$\Delta T=0.3\,T_{c}$~\cite{Shuryak1996}.
At small $T\leq T_{c}$ the solid line corresponds to  $M_{el}(0,T)/M_{el}(0,0)=\bar{\rho}^{'}(T)/\bar{\rho}(T)=(1-1/6\,\pi^{2}{\bar{\rho}_{0}}^{2}T^{2})$.
Dashed line  corresponds to the full suppression at the whole region
of $T$ ($A_{N_{c}}={\rm const}$ in \ref{Phi}). In both plots we
use for phenomenological estimates $M_{el}(0,0)=362\,{\rm MeV}$,
 $\bar{\rho}(0)=1/3\,{\rm fm}$ and $n(0)=1\,{\rm fm}^{-4}$. Right
plot: form-factor of the dynamical mass $F(q,0)$, Eq.~~(\ref{M01}). }
\label{Melfig} 
\end{figure}

We expect that the the dominant contribution to $\Pi_{\nu\mu}$ will
come from the large-distance asymptotics of the matrix elements of
$S_{\nu\mu}^{I}-{S^{0}}_{\nu\mu}$.  In coordinate space, comparing
the effects from $i\partial_{\mu}$ and from multiplication by $A_{\mu}^{I}$
in~(\ref{SI}), we conclude  that the dominant asymptotic contribution
to  $S_{\nu\mu}^{I}-{S^{0}}_{\nu\mu}$ in Eq. (\ref{barS-}) {comes}
from the term
\[
p_{\rho}({\rm {the\,\,most\,\,slowly\,\,decreasing}\,\,part\,\,of}\,\,(\Delta_{I}-\Delta_{0})\Delta_{0}+\Delta_{0}(\Delta_{I}-\Delta_{0}))p_{\sigma}
\]
\[
p_{\rho}((\Delta_{I}-\Delta_{0})\Delta_{0}+\Delta_{0}(\Delta_{I}-\Delta_{0}))p_{\sigma}
\]
and only this term will contribute to $\Pi_{\nu\mu}$. So, the Eq.~(\ref{barS-}) reduces to  
\begin{eqnarray}
\Pi_{\mu\nu}=2Np^{2}\left(\Delta_{I,0}-\Delta_{0}\right)\left(p^{2}\delta_{\mu\nu}-(1-1/\xi)p_{\mu}p_{\nu}\right)\label{Pi-}
\end{eqnarray}
By definition the square of \char`\"{}electric\char`\"{} gluon mass
$M_{el}^{2}(|\vec{k}|,T)$ is related to $\Pi_{\mu\nu}$ as $M_{el}^{2}(|\vec{k}|,T)=\Pi_{44}(\vec{k},k_{4}=0)$.
Comparing it with Eq. ~(\ref{Ms0}), we conclude that $M_{el}^{2}(|\vec{k}|,T)=2M_{s}^{2}(|\vec{k}|,T),$
is gauge invariant ($\xi$ independent) and its $T$ and $q$ dependencies
are represented by Fig.\ref{Melfig}. It is obvious that $M_{el}^{2}(|\vec{k}|,T=0)=M_{g}(|\vec{k}|)$,
where the gauge invariant dynamical gluon mass $M_{g}$ was obtained before~\cite{MO2018}.
Using the phenomenological values of $\bar{\rho}$ and $n$ at $T=0$,
we obtain $M_{el}(0,0)=362\,{\rm MeV}$. 

\begin{figure}[h]
\includegraphics[scale=0.6]{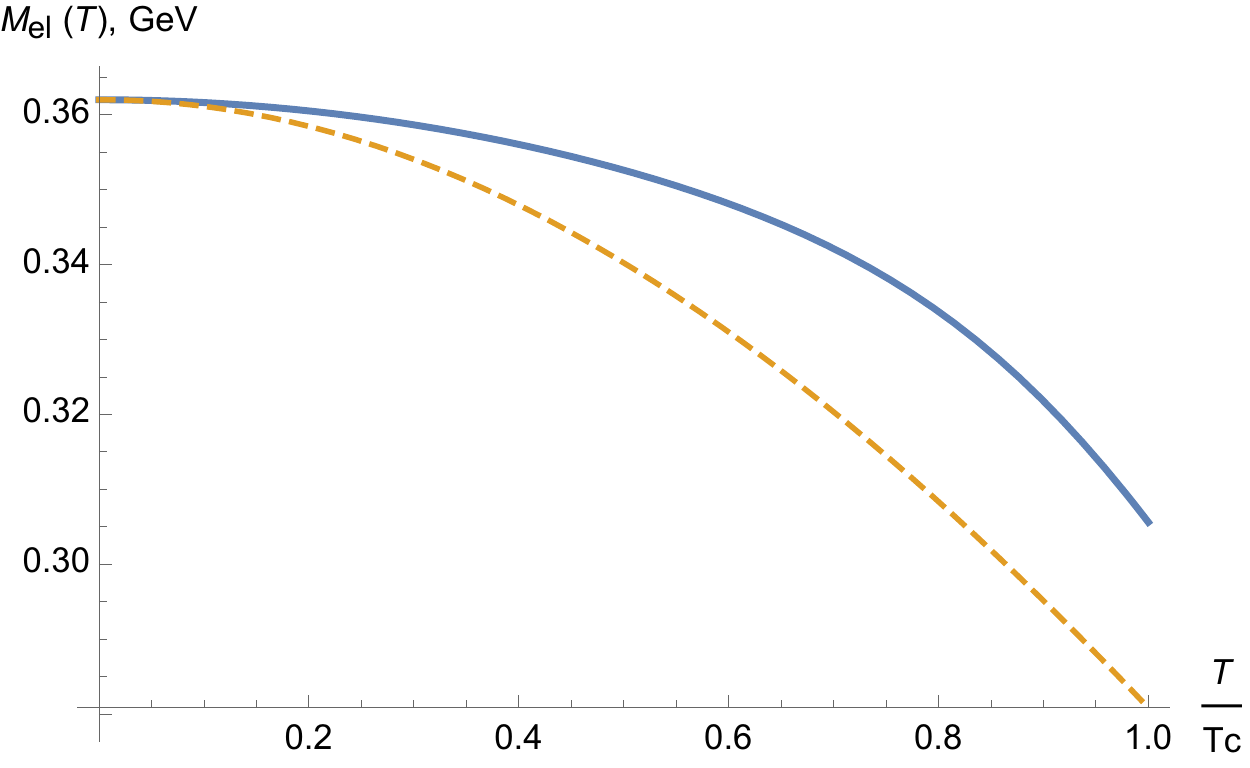}\includegraphics[scale=0.88]{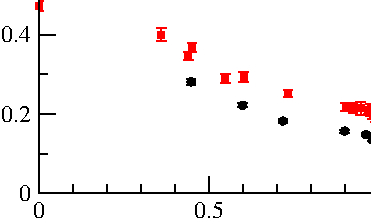}
\caption{Comparison of \char`\"{}electric\char`\"{} gluon dynamical masses
from ILM and from lattice measurements. \textbf{Left:} $T$-dependencies
of $M_{el}(0,T)$ in the region~$0<T<T_{c}$ within ILM (all definitions
are the same as in Fig.~\ref{Melfig}). \textbf{Right:} Lattice measurements
results for the same quantity taken at the scale $2\,{\rm GeV}$~\cite{Silva,Maas}
(Right plot is a part of Fig.12 from~\cite{Silva}. See the caption
of this Fig.).}
\label{Mel-lat} 
\end{figure}

From Fig.~\ref{Mel-lat} we see that the most natural explanation
of non-zero \char`\"{}electric\char`\"{} gluon dynamical mass at $T<T_{c}$
region seen in lattice measurements~\cite{Silva,Maas} is given by
ILM, since ILM is able at least qualitatively reproduce its value
at $T=0$ and its $T$ dependencies.

\section{Summary and Discussion}

\label{sec:Conclusion}In this paper we extended the calculations
of the dynamical gluon mass in ILM~\cite{MO2018} to non-zero temperature
and studied the so-called \char`\"{}electric\char`\"{} gluon mass
$M_{el}(q,T)$, which corresponds to $\Pi_{44}$-component of polarization
operator. We also analyzed the temperature $T$ dependence of the
main parameters of the ILM, the average instanton size $\bar{\rho}(T)$
and instanton density $n(T)$. We found that they are homogeneously
decreasing functions of temperature due to influence of thermal gluon
fluctuations~\cite{DM1988}. Our findings agree with lattice investigations
~\cite{Chu1995}, which demonstrated that $\bar{\rho}(T),\,\,n(T)$
are decreasing rapidly for $T\geq T_{c}$, where $T_{c}$ is the critical
temperature. For temperatures below the critical temperature $T_{c}$,
these functions are almost constant, and we took into account this
scenario by neglecting the contributions of thermal gluon fluctuations
at low temperature $T\leq T_{c}$~\cite{Shuryak1996}. The comparison
of both of these scenarios is presented at the Fig.~\ref{Shuryak1996}.

In order to find gluon propagator in the ILM background field at nonzero
temperature $T\neq0$, we have solved the gluon zero-modes problem
and averaged full gluon propagator over collective coordinates of
all instantons. This was done in the framework developed in our previous
paper~\cite{MO2018} and extended to non-zero temperature case. First,
we evaluated the \char`\"{}electric\char`\"{} color octet scalar
dynamical mass $M_{s}(q,\,T)$ as a function of the three-momentum
$\vec{q}$ and temperature $T$. The solution of zero-modes problem
yields $M_{el}^{2}(q,T)=2M_{s}^{2}(q,T)$, which allowed us to relate. 
The final results for the \char`\"{}electric\char`\"{} gluon dynamical
mass $M_{el}(q,T)$ are presented in the Fig.\ref{Melfig}.

It is interesting to compare our result for the dynamical \char`\"{}electric\char`\"{}
gluon mass $M_{el}$ with the result of lattice calculations (see
Fig.~\ref{Mel-lat}), which observed that $M_{el}(0,T)$ is a decreasing
function of $T$ for $T\leq T_{c}$ in the correspondence with ILM,
and is an increasing function of $T$ above the confinement-deconfinement
phase transition~\cite{Silva,Maas}. The grows of $M_{el}(0,T)$
for $T\geq T_{c}$ may be explained by perturbative thermal gluon
correction and is expected to have an almost linear functional dependence
$M_{pert,el}(0,T)\sim T$. Since thermal gluons are incorporated in
our framework, probably it is easy to reproduce within ILM model the
lattice measurements of the dynamical \char`\"{}electric\char`\"{}
gluon mass in whole region of temperature.

We assume to apply our result to the calculations of temperature dependence
of the heavy quarkonium properties.
\vskip 0.5cm
\textbf{Acknowledgments}

M.M. is thankful to Marat Siddikov and Boris Kopeliovich for the useful
and helpful communications. This work is partially supported by Fondecyt
(Chile) grant 1140377 and Uz grant OT-F2-10.

%\begin{thebibliography}{10}

\end{document}